\documentclass[aps,prd,showpacs,letterpaper,
 superscriptaddress,preprintnumbers,10pt]{revtex4}
\usepackage{epsfig,graphicx,bm}
\begin{document}
\newcommand{\be}{\begin{equation}}
\newcommand{\ee}{\end{equation}}
\newcommand{\bea}{\begin{eqnarray}}
\newcommand{\eea}{\end{eqnarray}}
\newcommand{\st}{{\scriptscriptstyle T}}
\newcommand{\xbj}{x_{\scriptscriptstyle B}}
\newcommand{\zh}{z_h}
\newcommand{\bfk}{\mbox{\boldmath $k$}}
\newcommand{\bfq}{\mbox{\boldmath $q$}}
\newcommand{\pup}{p^\uparrow}
\newcommand{\pdown}{p^\downarrow}
\newcommand{\qup}{q^\uparrow}
\newcommand{\qdown}{q^\downarrow}
\def\slash{\rlap{/}}
\newcommand{\bfp}{\mbox{\boldmath $p$}}
\newcommand{\bfP}{\mbox{\boldmath $P$}}
\newcommand{\Lup}{\Lambda^\uparrow}
\newcommand{\Ldown}{\Lambda^\downarrow}
\newcommand{\Aup}{A^\uparrow}
\newcommand{\hup}{h^\uparrow}
\newcommand{\hdown}{h^\downarrow}
\newcommand{\nd}{\noindent}
\newcommand{\la}{\lambda}
\def\lsim{\mathrel{\rlap{\lower4pt\hbox{\hskip1pt$\sim$}}\raise1pt\hbox{$<$}}}
\def\gsim{\mathrel{\rlap{\lower4pt\hbox{\hskip1pt$\sim$}}\raise1pt\hbox{$>$}}}
\def\nostrocostruttino#1\over#2{\mathrel{\mathop{\kern 0pt \rlap
{\hbox{$#1$}}} \hbox{\kern-.135em $#2$}}}
\def\sumint{\nostrocostruttino \sum \over {\displaystyle\int}}
\newcommand{\NP}[1]{{\it Nucl.\ Phys.}\ {\bf #1}}
\newcommand{\ZP}[1]{{\it Z.\ Phys.}\ {\bf #1}}
\newcommand{\PL}[1]{{\it Phys.\ Lett.}\ {\bf #1}}
\newcommand{\PR}[1]{{\it Phys.\ Rev.}\ {\bf #1}}
\newcommand{\PRL}[1]{{\it Phys.\ Rev.\ Lett.}\ {\bf #1}}
\newcommand{\MPL}[1]{{\it Mod.\ Phys.\ Lett.}\ {\bf #1}}
\newcommand{\SNP}[1]{{\it Sov.\ J.\ Nucl.\ Phys.}\ {\bf #1}}
\newcommand{\EPJ}[1]{{\it Eur.\ Phys.\ J.}\ {\bf #1}}
\newcommand{\IJMP}[1]{{\it Int.\ J.\ Mod.\ Phys.}\ {\bf #1}}
\newcommand{\simorder}{\raisebox{-4pt}{$\, \stackrel{\textstyle >}{\sim} \,$}}
\newcommand{\simordertwo}{\raisebox{-4pt}{$\, \stackrel{\textstyle <}{\sim}\,$}}

\title{Constraints on the gluon Sivers distribution via transverse single spin
asymmetries \\at midrapidity in $\bm{\pup \!\! p \to \pi^0 \, X}$ processes at BNL RHIC}
\author{M.~Anselmino}
 \email{mauro.anselmino@to.infn.it}
 \affiliation{Dipartimento di Fisica Teorica, Universit\`a di Torino and
              Istituto Nazionale di Fisica Nucleare, Sezione di Torino, \\
              Via P. Giuria 1, I-10125 Torino, Italy}
\author{U.~D'Alesio}
 \email{umberto.dalesio@ca.infn.it}
\author{S.~Melis}
 \email{stefano.melis@ca.infn.it}
\author{F.~Murgia}
 \email{francesco.murgia@ca.infn.it}
 \affiliation{Istituto Nazionale di Fisica Nucleare, Sezione di Cagliari and
              Dipartimento di Fisica, Universit\`a di Cagliari,\\
              C.P. 170, I-09042 Monserrato (CA), Italy}
\date{\today}

\begin{abstract}
We consider the recent RHIC data on the transverse single spin asymmetry (SSA) $A_N$,
measured in $\pup p \to \pi^0 \, X$ processes at mid-rapidity by the PHENIX collaboration.
The measurement is consistent with a vanishing SSA. We analyze this experimental information
within a hard scattering approach based on a generalized QCD factorization scheme, with
unintegrated, transverse momentum dependent (TMD), parton distribution and fragmentation
functions. It turns out that, in the kinematical region of the data, only the gluon Sivers
effect could give a large contribution to $A_N$; its vanishing value is thus an indication
about the possible size of the gluon Sivers function (GSF). Approximate upper limits on its
magnitude are derived. Additional constraints obtained combining available parameterizations
of the quark Sivers function and the Burkardt sum rule (BSR) for the Sivers distributions are
also discussed.
\end{abstract}


\maketitle

\section{Introduction and approach}
\label{intr}

We have recently discussed a hard scattering approach to hadronic
interactions, based on the assumption of a generalized QCD factorization
scheme which involves unintegrated TMD parton distribution and fragmentation
functions; the partonic intrinsic motions are also fully taken into account
in the elementary perturbative QCD (pQCD) processes \cite{fu,noi-1,noi-2}.
This scheme has been applied to the computation of unpolarized cross sections
for single-inclusive particle production, $p \, p \to h \, X$, at high energy
and (moderately) large $p_T$ \cite{fu}, and to the computation of the
transverse SSA
\be\label{defan}
A_N = \frac{d\sigma^\uparrow - d\sigma^\downarrow}
           {d\sigma^\uparrow + d\sigma^\downarrow}
\ee
in $\pup p \to h \, X$ processes \cite{fu,noi-1,noi-2,noi-3}.

Transverse single spin asymmetries can originate, even with a short distance
helicity conserving pQCD dynamics, from spin-$\bfk_\perp$ correlations in the
soft components of the hadronic process, the distribution and fragmentation
functions. There could be many such correlations. However, the study of SSA
based on the assumption of a generalized factorization scheme
\cite{noi-1,noi-2} shows that the correct treatment of the elementary pQCD
dynamics, with all phases related to the non-collinear and non-planar partonic
configurations, leads to a strong suppression of all contributions, with the
exception of the Sivers \cite{siv}, and, to a lesser extent, the Collins
\cite{col} mechanisms.

 The issue of the validity of the factorization scheme with unintegrated
 partonic distribution and fragmentation functions is still an open one.
 Such a scheme has been shown to hold for Semi-Inclusive Deep Inelastic Scattering (SIDIS)
 and Drell-Yan processes \cite{cs,jmy,cm}, while it is not yet clear whether
 or not it holds for inclusive one-particle production in $p\,p$
 processes;
 in such a case it is difficult to account
 for the gauge links necessary to ensure the gauge invariance of the TMD
 distribution functions \cite{bhs,col02,bjy,bmp,bbmp}. At this stage we
 consider our factorized description of $p \, p \to \pi \, X$ processes as a
 phenomenological model based on a natural extension of the usual collinear
 approach for the same process, and of the factorized scheme with unintegrated
 partonic distributions proven for SIDIS and Drell-Yan processes.

The Sivers function has recently received a lot of attention: data on
azimuthal SSA in semi-inclusive deep inelastic scattering (SIDIS) processes
from the HERMES collaboration at DESY~\cite{her} and from the COMPASS
collaboration at CERN~\cite{com} have allowed, for the first time, a direct
extraction of the Sivers functions for $u$ and $d$ quarks inside a proton
\cite{noi-4,vog,efr,goe,como}. Similarly, $u$ and $d$ Sivers functions have been
extracted from purely hadronic processes \cite{fu}, in qualitative agreement
with the SIDIS results. Also the possibility of accessing the gluon Sivers
function has been investigated \cite{noi-3,m-r,b-v,sof}. Although one might
expect that spin-$\bfk_\perp$ correlations are stronger for valence quarks --
as the large $x_F$ data from the FNAL-E704 collaboration~\cite{e704} and the
STAR collaboration at RHIC-BNL~\cite{star} seem to indicate -- one knows that
gluons play a dominant role in many high energy hadronic processes; it
would be very interesting to see whether or not the gluon density inside a
transversely polarized proton depends on the intrinsic motion.

We address here the issue of the largely unknown gluon Sivers function,
$\Delta^N\!\hat f_{g/\pup}(x,k_\perp)$, and its possible contribution to
the SSA $A_N$ for the $\pup p \to \pi^0 \, X$ process, in the framework of
the generalized factorization scheme, with pQCD elementary dynamics, of
Refs.~\cite{fu,noi-1,noi-2}. In this approach, the general structure of the
cross section for the polarized hadronic process
$(A,S_A) + (B,S_B) \to C + X$, is given by \cite{noi-2}
\bea
\frac{E_C \, d\sigma^{(A,S_A) + (B,S_B) \to C + X}}
{d^{3} \bfp_C} = \!\!\!\!\! \sum_{a,b,c,d, \{\la\}}
\!\!\!\!\!\!\!\!\!\!\!\! && \int \frac{dx_a \,
dx_b \, dz}{16 \pi^2 x_a x_b z^2  s} \;
d^2 \bfk_{\perp a} \, d^2 \bfk_{\perp b}\, d^3 \bfk_{\perp C}\,
\delta(\bm{k}_{\perp C} \cdot \hat{\bm{p}}_c) \, J(k_{\perp C})
\nonumber \\
\hskip-12pt &\times& \rho_{\la^{\,}_a,
\la^{\prime}_a}^{a/A,S_A} \, \hat f_{a/A,S_A}(x_a,\bfk_{\perp a})
\> \rho_{\la^{\,}_b, \la^{\prime}_b}^{b/B,S_B} \,
\hat f_{b/B,S_B}(x_b,\bfk_{\perp b}) \label{gen1} \\
\hskip-12pt &\times&
\hat M_{\la^{\,}_c, \la^{\,}_d; \la^{\,}_a, \la^{\,}_b} \,
\hat M^*_{\la^{\prime}_c, \la^{\,}_d; \la^{\prime}_a,
\la^{\prime}_b} \> \delta(\hat s + \hat t + \hat u) \> \hat
D^{\la^{\,}_C,\la^{\,}_C}_{\la^{\,}_c,\la^{\prime}_c}(z,\bfk_{\perp C})
\>, \nonumber
\eea
where all parton intrinsic motions are fully taken into account, both
in the soft, non perturbative components and in the hard, pQCD interactions.

In Ref.~\cite{noi-2} the structure of Eq.~(\ref{gen1}) was extensively discussed; its main
features are the appearance of several spin and $\bfk_\perp$ dependent distribution and
fragmentation functions (with a partonic interpretation) and the non-collinear partonic
configurations which lead to many $\bfk_\perp$ dependent phases. As a consequence, it was
explicitly shown how the integration over the parton intrinsic momenta leads to strong
suppressions of most contributions to the unpolarized cross sections and to the transverse
single spin asymmetry $A_N$. The only sizeable contributions to the latter, in the
kinematical region (large positive $x_F$) of the E704 \cite{e704} and STAR data \cite{star}
come from the Sivers and, less importantly, from the Collins mechanisms. The dominance of the
Sivers effect is even more pronounced in other $x_F$ ranges, namely at $x_F \leq 0$.

In Ref.~\cite{noi-2} also the flavour decomposition of the Sivers effect was
performed; while the quark contribution is totally dominant at large and
positive $x_F$ values (for polarized protons moving along the positive
$Z$-axis), the gluon contribution may be sizeable at $x_F \simeq 0$, the
mid-rapidity region. This can be easily understood, since $x_{a}^{\rm{min}}$
(the lowest kinematically accessible value of the light-cone momentum fraction of parton $a$ inside the transversely polarized initial proton) must be
larger than $x_F$; $A_N$ at large $x_F$ is then mainly driven by valence
quark properties. Indeed, an analysis of the E704 data allowed a first extraction of the Sivers functions for $u$ and $d$ quarks \cite{fu}. In
principle, also inclusive production in the negative $x_F$ region might be
sensitive to small $x_a$ gluons inside the polarized proton (being hit by
large $x_b$ partons inside the unpolarized one). We shall further comment
on this point after Eq.~(\ref{unpol}).

Data in the mid-rapidity region are available from the E704~\cite{e704y0}
and PHENIX~\cite{phe} experiments. The kinematical regime corresponding to
negative values of $x_F$ has been covered by the STAR
collaboration~\cite{star06}. Preliminary results  for charged pions are also
available from the RHIC-BRAHMS experiment~\cite{brahms}.

In these kinematical regions $A_N(\pup p\to\pi\,X)$ is largely dominated by the Sivers effect
alone, and Eq.~(\ref{gen1}) gives \cite{fu}:
\bea \!\!\!\!\! && \frac{E_\pi \, d\sigma^\uparrow}{d^3\bfp_\pi} -
\frac{E_\pi \, d\sigma^\downarrow}{d^3\bfp_\pi} \simeq \sum_{a,b,c,d} \int \frac{dx_a \, dx_b \, dz}{\pi \, x_a \, x_b \, z^2 \, s} \; d^2\bfk_{\perp a} \, d^2\bfk_{\perp b} \,
\,d^3\bfk_{\perp \pi} \, \delta(\bfk_{\perp \pi}\cdot \hat{\bfp}_c) \, J(k_{\perp \pi})
\nonumber \\ \!\!\!\!\! &\times& \Delta \hat f_{a/\pup}(x_a, \bfk_{\perp a})
\> \hat f_{b/p}(x_b, k_{\perp b})\,\hat s^2 \>\frac{d\hat\sigma^{ab \to cd}}
{d\hat t}(x_a,x_b, \hat s, \hat t, \hat u) \> \delta(\hat s + \hat t + \hat u) \> \hat D_{\pi/c}(z, k_{\perp \pi}) \>,
\label{sivgen} \eea
where ($M$ denotes the proton mass)
\be
\Delta \hat f_{a/\pup}\,(x_a, \bfk_{\perp a}) \equiv
\hat f_{a/\pup}\,(x_a, \bfk_{\perp a}) - \hat f_{a/p^\downarrow}\,
(x_a, \bfk_{\perp a})
\label{defsiv}
= \Delta^N \hat f_{a/\pup}\,(x_a, k_{\perp a}) \> \cos\phi_a
=  -2 \, \frac{k_{\perp a}}{M} \, f_{1T}^\perp (x_a, k_{\perp a}) \>
\cos\phi_a \>.
\ee
$\Delta^N \hat f_{a/\pup}(x_a, k_{\perp a})$
[or $f_{1T}^\perp (x_a, k_{\perp a})]$ is referred to as the Sivers
distribution function of parton $a$ inside a transversely polarized proton.
$\phi_a$ is the azimuthal angle of the intrinsic transverse momentum
$\bfk_{\perp a}$ of parton $a$. We follow here the notations and kinematical
conventions of Ref.~\cite{noi-2}, with the polarized proton moving
along the positive $Z$-axis, in the $pp$ c.m.~frame; the observed pion is
produced in the $XZ$ plane, with positive $X$ values; spin $\uparrow$ and
$\downarrow$ are respectively along the positive and negative $Y$-axis.

Eq.~(\ref{sivgen}) gives the numerator of $A_N$, Eq.~(1); the denominator is
just twice the unpolarized cross section, which is given by
\bea
\!\!\!\!\! && \frac{E_\pi \, d\sigma^{unp}}{d^3\bfp_\pi}
\simeq \sum_{a,b,c,d} \int
\frac{dx_a \, dx_b \, dz}{\pi \, x_a \, x_b \, z^2 \, s} \;
d^2\bfk_{\perp a} \, d^2\bfk_{\perp b} \, \,d^3\bfk_{\perp \pi} \,
\delta(\bfk_{\perp \pi}\cdot \hat{\bfp}_c) \, J(k_{\perp \pi})
\nonumber \\ \!\!\!\!\!
&\times& \hat f_{a/p}(x_a, k_{\perp a}) \>
\hat f_{b/p}(x_b, k_{\perp b}) \, \hat s^2 \>
\frac{d\hat\sigma^{ab \to cd}}{d\hat t}(x_a,x_b, \hat s, \hat t, \hat u) \>
\delta(\hat s + \hat t + \hat u) \> \hat D_{\pi/c}(z, k_{\perp \pi})
\>. \label{unpol}
\eea

The azimuthal phase factor $\cos\phi_a$ appearing in the numerator of $A_N$,
Eqs.~(\ref{sivgen}) and (\ref{defsiv}), plays a crucial role and deserves a comment. The only
other term depending on $\phi_a$ in Eq.~(\ref{sivgen}) is the partonic cross section, in
particular via the corresponding Mandelstam variable $\hat{t}$. At large positive $x_F$ and
moderately large $p_T$ the (average) values of $\hat{t}$ are relatively small. Therefore, the
(dominant) $\hat{t}$-channel contributions, proportional to $1/\hat t^{\,2}$, depend sizeably
on $\phi_a$, so that the $d^2\bm{k}_{\perp a} \, \cos\phi_a / \hat t^{\,2}$ integration in
Eq.~(\ref{sivgen}) does not necessarily suppress $A_N$. Instead, for negative values of
$x_F$, all partonic Mandelstam variables are much less dependent on $\phi_a$, so that one is
roughly left with the $d^2\bm{k}_{\perp a}\,\cos\phi_a$ integration alone, which cancels the
potentially large Sivers contribution. As a consequence, the possibility of gaining
information on the gluon Sivers distribution from the recent STAR and BRAHMS data at negative
values of $x_F$ is frustrated. Notice that, due to the much lower values of $\sqrt{s}$
involved, this suppression caused by the $\cos\phi_a$ dependence would be much less effective
for the kinematical regimes of E704 and of the proposed PAX~\cite{pax} experiments
\cite{noi-2}. However, E704 never measured $A_N$ for negative $x_F$ values, while the PAX
experiment is still in the proposal and planning stage.

Similar considerations lead to a strong suppression of the gluon Sivers
contribution to the SSA for $p^{\uparrow} p\to\gamma\,X$ processes at STAR,
in the negative $x_F$ range and at a (pseudo)rapidity $\eta \simeq -4$
\cite{unp}.

The same arguments do not apply to inclusive hadronic processes at mid-rapidity and
moderately large $p_T$ values, for which data from PHENIX~\cite{phe} are already available,
for neutral pions and charged hadron production. For these processes, the gluon contribution
is dominant and the Sivers effect can survive the phase integration. In the next section, we
shall therefore consider in detail this case, aiming at a possible derivation, from data, of
useful direct constraints on the gluon Sivers function.

\subsection{Constraints from the Burkardt sum rule}

Indirect constraints on the GSF could also be obtained from a sum rule for the Sivers distribution recently derived by Burkardt~\cite{burk}. The BSR states
that the total (integrated over $x$ and $\bm{k}_\perp$) transverse momentum of all partons (quarks, antiquarks and gluons) in a transversely polarized proton must be zero,
\begin{equation}
 \langle\bm{k}_{\perp}\rangle =
 \sum_a \;\langle\bm{k}_{\perp}\rangle_a =
 \int\!dx \int d^{\,2}\bm{k}_\perp \,\bm{k}_\perp
 \sum_a \Delta \hat{f}_{a/p^{\uparrow}}(x,\bm{k}_\perp) = 0 \, .
 \label{rule}
\end{equation}

Naively, at the partonic level considered in our approach, this sum rule simply corresponds
to total (transverse) momentum conservation inside a transversely polarized proton, since the
unintegrated distribution function
\begin{equation}
\hat f_{a/p^{\uparrow}}\,(x_a, \bfk_{\perp a}) =
\hat f_{a/p}\,(x_a, k_{\perp a}) + \frac 12 \,
\Delta \hat f_{a/p^{\uparrow}}\,(x_a, \bfk_{\perp a}) \>,
\end{equation}
introduced in Eq.~(\ref{defsiv}), has a clear probabilistic interpretation.
On the other hand, it is well known \cite{bhs,col02} that a non-vanishing
Sivers effect requires initial/final state interactions, which might spoil
the simple partonic interpretation. The Burkardt sum rule ensures the
non-trivial result that momentum conservation holds also in this situation.
The validity of the BSR has been explicitly verified
in a diquark spectator model calculation of the Sivers distribution in
Ref.~\cite{goe-2}. In this paper, we consider the BSR as an additional
theoretical tool in order to obtain indirect information on the gluon Sivers
distribution, once the quark distributions are known.

We notice, however, that a strict use of the BSR requires integration over the full $x$ range, including the poorly known small $x$ region, of the single Sivers functions, which might even result in divergences. Therefore, in the following we shall simply check whether or not the parameterizations of the
Sivers functions, which will turn out from our phenomenological analysis of the data within the theoretical approach of Refs.~\cite{fu,noi-1,noi-2}, fulfill
the BSR in the limited $x$ range considered, without extrapolating them to
very low $x$ values.

In the next section we present our results on the GSF, based on the
PHENIX~\cite{phe} experimental data on $A_N(p^\uparrow p\to\pi^0 \, X)$.
Further comments and conclusions are given in the last section.

\section{Phenomenology of the gluon Sivers function}
\label{ris}

We consider the PHENIX data \cite{phe} on $A_N$ for the $\pup p \to
\pi^0 \, X$ process at RHIC, at $\sqrt s$ = 200 GeV, with $p_T$ ranging from
1.0 to 5.0 GeV/$c$ and mid-rapidity values, $|\eta| < 0.35$. In this
kinematical regime, at the lowest $p_T$ values, $x_{a}^{\rm{min}}$ can be as
small as 0.005. Therefore, partonic channels involving a gluon in the transversely polarized initial proton dominate over those involving
a quark. This gluon dominance, together with the (almost) vanishing of all
possible contributions to $A_N$ other than the Sivers effect, allow to interpret the data -- showing tiny values of $A_N$ -- in terms of useful
constraints (upper bounds) on the gluon Sivers function, as it will be shown.
As $p_T$ grows, $x_{a}^{\rm{min}}$ increases and the dominance of the gluon
channels becomes less prominent.

Another set of data, for comparable rapidity and $p_T$ ranges, has
been collected several years ago by the E704 Collaboration \cite{e704y0},
at a lower energy, $\sqrt{s}\simeq 20$ GeV. In this case, however, even at
the smallest $p_T$ values, $x_{a}^{\rm{min}}$ remains large enough so that
gluon channels are not dominating; a possible mixing with quark initiated
contributions then prevents to get clean constraints on the GSF from E704
lower energy data.

Before analyzing in detail the reliable (for our purpose) PHENIX data, let us summarize what
we know so far about the Sivers functions. As already discussed, the E704 and STAR
collaborations have measured large SSA for the $p^\uparrow p\to \pi X$ process at large
positive $x_F$ and moderate $p_T$ values. In this kinematical regime, since $x_{a}^{\rm{min}}
> x_F$, gluon and sea-quark contributions should be negligible. Indeed, in Ref.~\cite{fu} we
have shown that reasonable fits to the SSA can be obtained by using valence-like Sivers
functions for $u$ and $d$ quarks, which turn out to have opposite signs. In addition, fits to
the weighted azimuthal asymmetries measured for pion production in SIDIS with a transversely
polarized target \cite{her,com}, lead to independent comparable parameterizations of the $u$
and $d$ Sivers distributions \cite{noi-4,vog,goe,como}.

Notice that in all these phenomenological analyses the Sivers functions for gluons and
sea-quarks have been assumed to be vanishing. That is a reasonable assumption, considering
the kinematical region of the available data, and the lowest order decoupling of gluons in
SIDIS. A natural question arising at this point is the following: how do the valence $u$ and
$d$ Sivers functions alone, so far extracted, perform with the mid-rapidity PHENIX and E704
data? The answer is that they predict an almost vanishing SSA, compatible with both sets of
data (see Fig.~\ref{an}, for the PHENIX results). Not only, but their
parameterizations~\cite{fu,noi-4} are also compatible with the Burkardt sum rule,
Eq.~(\ref{rule}), yielding $\langle\bm{k}_{\perp}\rangle\simeq 0$ within a 10\% accuracy. It
looks like there is no need to introduce other contributions to the Sivers effect in addition
to that coming from valence quarks.

Of course, this cannot be a definite, although simple and appealing,
conclusion, as the data used for the extraction of the $u$ and $d$ quark
Sivers functions are insensitive to the small $x$ region. A large gluon
Sivers function would not affect the analysis of the SIDIS, E704 and STAR
data at large positive $x_F$; however, it would strongly affect the
description of the mid-rapidity PHENIX data. The real question is now:
how much does the small value of $A_N$ \textit{measured} by PHENIX suppress
the gluon Sivers function?

In order to answer this question we compute $A_N$ according to Eqs. (\ref{defan}) and
(\ref{sivgen})-(\ref{unpol}), imposing different conditions on the gluon Sivers functions,
trying to understand what is the maximum value of $|\Delta^N \hat
f_{g/\pup}\,(x,k_{\perp})|/2\hat f_{g/p}\,(x, k_{\perp})$ allowed by the PHENIX data. We
follow Ref.~\cite{fu}, adopting the same factorized gaussian $k_\perp$ dependence for
distribution and fragmentation functions and the same parameterization for the Sivers
distributions, which are related to the unpolarized parton densities; the latter are given by
the MRST01 set \cite{MRST01}, and the fragmentation functions by the KKP set \cite{KKP}. The
valence $u$ and $d$ quark Sivers functions are the same as extracted from E704 data in Ref.
\cite{fu}.

Our results are shown in Figs.~\ref{an} and \ref{bound} and require some comments.

\noindent $\bullet$ The thin, red, solid line in Fig.~\ref{an}, is the result of computing
$A_N$ using only the valence $u$ and $d$ Sivers functions, as extracted in Ref. \cite{fu}.\\
$\bullet$ The cyan, dot-dashed curve in Fig.~\ref{an} corresponds to the extreme case of the
largest (in magnitude) gluon Sivers distribution, obtained by saturating the natural
positivity bound, see Eq.~(\ref{defsiv}),
\begin{equation}
\Delta^N \hat f_{g/\pup}\,(x,k_{\perp}) = -
2\hat f_{g/p}\,(x, k_{\perp}) \>. \label{gsfmax}
\end{equation}
The sea-quark Sivers functions are assumed to vanish. This leads to a SSA definitely in
contradiction with the data. It also leads to a strong violation of the BSR. As expected,
from PHENIX data a much more stringent constraint than the simple positivity bound for the
(absolute value of the) gluon Sivers function can be obtained.
The maximized (in magnitude) GSF in
 Eq. (\ref{gsfmax}) has been chosen negative as the data hint at a possible
 small negative value of $A_N$. The opposite choice would lead to very similar
 conclusions, actually with an even stronger disagreement with data.\\
$\bullet$ The thick, red, solid curve in Fig.~\ref{an} has been obtained still assuming that
there is no sea-quark Sivers contribution, and looking for a parameterization of $\Delta^N
\hat f_{g/\pup}$ yielding values of $A_N$ falling, approximately, within one-sigma deviation
below the lowest $p_T$ data. This somehow corresponds to the largest (in magnitude)
acceptable gluon Sivers function, in the sense that any larger GSF would cause the SSA to lie
outside the error bars.

The corresponding $x$-dependent part of the GSF, normalized to its positivity bound,
$|\Delta^N f_{g/\pup}(x)|/ 2 f_{g/p}(x)$, is shown as the red, solid curve in
Fig.~\ref{bound}, as a function of $x$. PHENIX data on $A_N$ clearly impose a stringent upper
bound on the magnitude of the gluon Sivers function in the region of small $x$, where gluons
play a crucial role. The constraint is much less significant at larger $x$ values, as it is
natural: there cannot be any strong correlation between gluon distributions at large $x$
(where they are anyway negligible), and a physical observable, like $A_N$ at mid-rapidity,
which is mainly sensitive to small $x$ values.

The GSF obtained here, the red, solid curve of Fig.~\ref{bound}, leads, within the $x$ range
covered by the data, to a strong violation of the BSR. This could be avoided by imposing an
even smaller (in size) GSF, giving a strong role to the BSR.\\
$\bullet$ As a last attempt, we have released the assumption of vanishing sea-quark Sivers
distributions. With the aim of exploring how large a GSF can be, we have done that in an
extreme scenario, \textit{i.e.} assuming that all sea-quarks ($u_s$, $\bar{u}$, $d_s$, $\bar{d}$,
$s$, $\bar{s}$) have a non-vanishing positive Sivers function which saturates the positivity
bound [that is, $\Delta^N \hat f_{q_s/\pup}\,(x, k_{\perp}) \equiv 2\hat f_{q_s/p}\,(x,
k_{\perp})$]. Their contribution could then cancel the negative contribution to $A_N$ of a
possibly large GSF.

Again, we look for the largest negative GSF which, together with a positive maximized
sea-quark contribution, gives a SSA approximately lying within one-sigma standard deviation
below the data points, represented again by the thick, red, solid line of Fig.~\ref{an}. This
curve results now as the sum of the (maximized) sea and valence quark contribution (blue,
dotted curve) and of the new GSF (green, dashed curve).

The corresponding, normalized, new GSF is plotted as the green, dashed curve
in Fig.~\ref{bound}. It is the largest gluon Sivers function compatible with
PHENIX data: actually, it is strongly artificially enhanced by the extreme
(or rather, unrealistic) assumption about the opposite contribution from
saturated (and all summing up) sea-quark Sivers functions. Nevertheless, it
still indicates a rather modest GSF in the important small $x$
region.  Notice that the opposite choice, a negative contribution from the sea-quark
 Sivers distributions balancing a positive one from the GSF, would lead to an
 even smaller GSF.

Finally, we notice that, concerning the BSR, within the $x$ range covered by the data, the
(over)maximized sea-quark Sivers distributions give a positive contribution which strongly
suppresses the negative contribution of the GSF, so that in this scenario the BSR is
satisfied within a 10\% level accuracy.

\section{Comments and conclusions}

Unintegrated, TMD parton distribution and fragmentation functions may help in explaining
several puzzling measurements for spin observables in inclusive particle production at high
energy and moderately large $p_T$ (for hadronic collisions) or $Q^2$ (for SIDIS). In
particular, the quark Sivers distributions and the Collins fragmentation functions have
recently raised a lot of interest. Azimuthal and single spin asymmetries measured by several
experimental collaborations unambiguously indicate that these effects are sizeable, at least
in some kinematical regions.

A combined analysis of inclusive pion production in hadronic collisions and
in SIDIS has given useful information on the quark Sivers distributions in the
valence region. On the contrary, sea-quark and gluon Sivers functions are
largely unknown. In this paper, we have performed a phenomenological analysis
of the available data for the SSA $A_N(p^\uparrow p\to\pi X)$, in particular
regarding the recent PHENIX data obtained at mid-rapidity and moderate $p_T$
values; this SSA is dominated by the Sivers effect and explores the low $x$
region, resulting very sensitive to gluon contributions. To this end, we
have adopted the theoretical approach of Refs.~\cite{fu,noi-1,noi-2}.

A first result of our analysis is that all available data, including the PHENIX ones, are
compatible with valence-like quark Sivers distributions and vanishing sea-quark and gluon
contributions. The parameterizations required to reasonably reproduce the observed
asymmetries also fulfill, within a 10\% accuracy, the Burkardt sum rule. Since these
parameterizations have a valence-like nature, the sum rule can be checked both over the fully
integrated $x$ range and in the limited $x$ range effectively covered by data, with the same
conclusions.

\begin{center}
\begin{figure}[t]
\epsfig{file=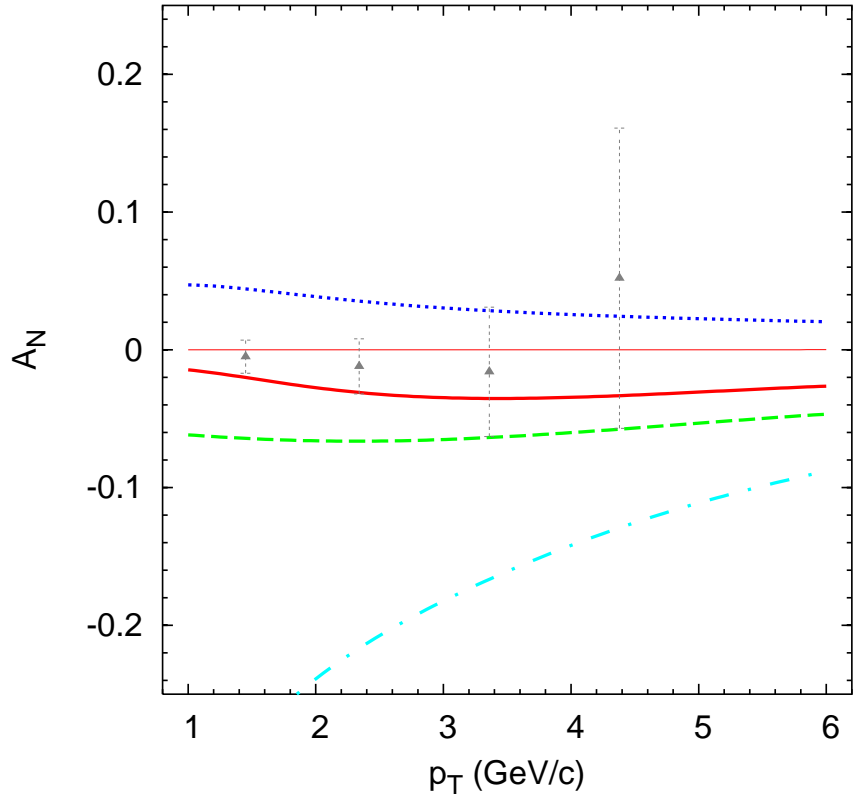,angle=-90,width=12.0truecm} \caption{The SSA $A_N$, computed
according to Eqs.~(\ref{defan}) and (\ref{sivgen})-(\ref{unpol}) of the text, and compared
with PHENIX data \cite{phe}, with different choices for the gluon and sea-quark Sivers
functions. The thin, red, solid line shows the results given by the valence $u$ and $d$
Sivers functions alone. The cyan, dot-dashed curve shows the contribution of the maximized
GSF alone, $\Delta^N \hat f_{g/\pup} = -2\hat f_{g/p}$. The thick, red, solid curve is
obtained by requesting results within one standard deviation from data in two different ways:
no sea-quark contribution and maximized sea-quark contribution. In the latter case the blue,
dotted curve shows the contribution of the sea (maximized) + valence quarks, while the green,
dashed curve shows the contribution of the GSF. The valence quark Sivers distributions are
taken from Ref.~\cite{fu}, while the unpolarized PDF's and FF's from Refs.~\cite{MRST01} and
\cite{KKP} respectively. Further detail can be found in the text.} \label{an}
\end{figure}
\end{center}

\begin{center}
\begin{figure}[t]
\epsfig{file=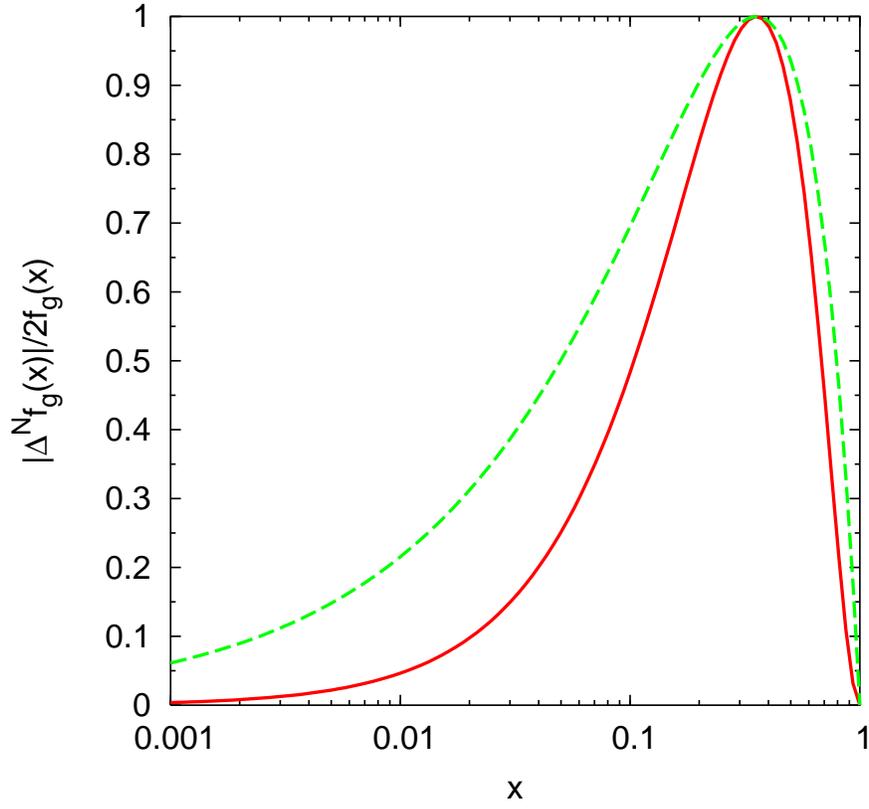,angle=-90,width=12.0truecm}
\caption{The value of the normalized GSF, $|\Delta^N f_{g/\pup}(x)|
/2 f_{g/p}(x)$, as obtained from fitting, within one standard
deviation, the PHENIX data on $A_N$. The red, solid curve corresponds to
the case of vanishing sea-quark contribution; the green, dashed curve shows
the GSF corresponding to the extreme scenario in which the sea-quark
contributions are all maximized and sum up to balance the negative GSF
contribution. Further detail can be found in the text.}
\label{bound}
\end{figure}
\end{center}

The main issue of this paper was, however, that of discussing to what extent the available
data can bind the magnitude of the gluon Sivers function, in particular in the small $x$
region, where gluons play an important role. Our analysis shows that the PHENIX data on $A_N$
are presently the only ones that allow to reach quantitative conclusions on the magnitude of
the GSF. The weakest upper bound (that is, the largest GSF), is obtained by balancing the
gluon contribution to $A_N$ with (over)maximized sea-quark contributions, opposite in sign.
The resulting GSF is represented by the green, dashed line of Fig.~\ref{bound}. This scenario
corresponds to a SSA lying approximately within one-sigma deviation below the data, as shown
by the thick, red, solid line of Fig.~\ref{an}. Even in this extreme case, the bound obtained
significantly reduces the magnitudes available from the simple positivity bound implicit in
the definition of the Sivers function, at least in the $x$ region where gluon contributions
are relevant.

Some comments are still in order:\\
1) Following Refs.~\cite{fu,noi-1,noi-2}, we have assumed for the TMD
functions a simple factorized expression, with a Gaussian $k_{\perp}$
dependence. The same $\langle k^2_{\perp}\rangle\simeq 0.64$ (GeV/$c)^2$ has
been used for all partons (quarks and gluons). It has been suggested that
gluons may have a larger $\langle k^2_{\perp}\rangle$ than quarks
\cite{werner}. We have found that, using smaller values of
$\langle k^2_{\perp}\rangle_q$ and, \textit{e.g.},
$\langle k^2_{\perp}\rangle_g \simeq 2\,\langle k^2_{\perp}\rangle_q$
would lead to a more stringent bound on the GSF.\\
2) In our calculations we have used the MRST01-LO set for distribution
functions and the KKP-LO set of fragmentation functions. Since low-$x$ values
are relevant in the kinematical configuration of RHIC experiments, one may
wonder whether the choice of unpolarized PDF's and FF's affects the bounds.
We have checked that there is indeed some residual dependence of this type.
As an example, using the CTEQ6 PDF's \cite{cteq}, which have a larger gluon
component than the MRST01 set, would lead to a slightly more stringent bound.
On the contrary, the use of the Kretzer FF set \cite{kre}, which has a smaller gluon
fragmentation component, would lead to a slightly less stringent bound.
These small changes are well within the overall uncertainty of our results
and do not alter our main conclusions.

Future measurements covering the negative $x_F$ range, but at smaller c.m. energies,
\textit{e.g.} at RHIC with $\sqrt{s}=63$ GeV or at the proposed PAX experiment at GSI, with
$\sqrt{s}=14$ GeV, will certainly help in clarifying the size and relevance of the GSF. SSA
measurements for inclusive photon production might also give useful information. A detailed
analysis of the phenomenological interest of these processes is in progress \cite{fsu}.

\acknowledgments

We acknowledge the support of the European Community--Research Infrastructure
Activity under the FP6 ``Structuring the European Research Area'' programme
(HadronPhysics, contract number RII3-CT-2004-506078); M.A.
acknowledges partial support by MIUR under contract 2004021808\_009.

\vskip 18pt
 {\bf Note added in proof}
 \vskip 6pt \noindent
 Just after submission of our paper an interesting work appeared \cite{bg} in
 which similar conclusions about the smallness of the gluon Sivers function
 are obtained from the smallness of the Sivers effect in SIDIS off a deuteron
 target.


\end{document}